# Research information in the light of artificial intelligence: quality and data ecologies

Otmane Azeroual [1], Tibor Koltay [2]

[1] German Centre for Higher Education Research and Science Studies (DZHW), Schützenstraße 6a, 10117 Berlin, Germany
[2] Eszterházy Károly Catholic University, HU-3300 Eger, Hungary

**Abstract**
This paper presents multi- and interdisciplinary approaches for finding the appropriate AI technologies for research information. Professional research information management (RIM) is becoming increasingly important as an expressly data-driven tool for researchers. It is not only the basis of scientific knowledge processes, but also related to other data.
A concept and a process model of the elementary phases from the start of the project to the ongoing operation of the AI methods in the RIM is presented, portraying the implementation of an AI project, meant to enable universities and research institutions to support their researchers in dealing with incorrect and incomplete research information, while it is being stored in their RIMs. Our aim is to show how research information harmonizes with the challenges of data literacy and data quality issues, related to AI, also wanting to underline that any project can be successful if the research institutions and various departments of universities, involved work together and appropriate support is offered to improve research information and data management.

## 1. Introduction

The amount of data, defined as a "reinterpretable representation of information in a formalized manner, suitable for communication, interpretation, or processing" [1] is constantly increasing in varied institutions.

Particularly affected is the amount of research information (such as publication data, personal data, project data, third-party funded data, etc.) in universities and research institutions. This means that research results can not only be verified and interpreted, but it must be understood how these results came about and how they can be used. As the preparation, utilization and preservation of a wide variety of research information has always been an important core task for these institutions and their libraries, as they can take over the organization of all information about the data stocks and their secure long-term archiving.

The usefulness of useful research information depends very much on the quality of the data, recorded there. Nowadays, the topic of data quality (DQ) is becoming therefore more and more important both in theory and practice. This is not surprising, since securing and improving it is playing an increasingly important role, especially in the course of rapidly growing data stocks and the increasing use of RIM. Data quality is defined as properties of data in relation to their ability to meet specified requirements [2,3]. To ensure a high level of DQ, scientifically proven methods and procedures are required.

Although the methods of controlling traditional data quality are based on user experience or pre-established business rules that limits performance in addition to a very time-consuming process and low accuracy [4,5], there seems to be a need for making use applying views that can be regarded as traditional and related to data ecologies in a wide sense.

Anyhow, providing research information alone does not add value to institutions. Only when research information is enriched by adding relevance and purpose to it for instance –through analysis and processing – information is created and this can then generate added value in the further course. In order to evaluate relevant research information, institutions must increasingly rely on new technologies such as artificial intelligence (AI). On the one hand, AI methods are used so that they can deal with today's complexity of data and information in an automated and self-learning manner. On the other hand, to make the data accessible for analysis and evaluation, we need to be familiar with methods for the collection, preparation and processing of large amounts of data. Such methods are therefore necessary today in many respects if we want to exploit the potential of data.

Beyond these introductory thoughts, the rest of our paper is structured as follows: (2) gives a view on data ecologies, then (3) deals with related work and the state of the art; (4) describes the AI methods to improve data quality, such as pattern recognition; (5) shows, how the AI methods can be comprehensively implemented in RIM and as a project at institutions. We offer a creative concept based on artificial intelligence algorithms to improve and manage data quality and informed decisions of research information in RIM. It is aimed at those, who are responsible for data management and quality assurance in research institutions (6) we discuss our results and (7) provide a summary of our paper.

## 2. Views on data ecologies

Information ecology examines the complex relationships between human beings in information places and spaces, manifest in the form of sense-making [6].

When defining data ecology, we can change the word "information" to "data". This means that due to the lack of strict lines of division between information and data, the concept of ecology can be applied to both. Moreover, we can also accept that data ecology can provide a basis for data literacy that examines "how people navigate and interpret data through interactions with tools and other people" [7].

### 2.1. Data ecology

One of the evident forms of data ecology is Research Data Management (RDM) and also addressed as Research Data Services (RDS). RDM can be defined as a "series of practices for dealing with resources," by involving "the broad stages of planning and preparing, conducting, archiving, publishing and disseminating, using a research lifecycle model" [8].

The existence of Data Management Plans (DMPs), often mandated by research funders, as well as the growth of these plans [9] shows that the landscape of research data is changing.

Data journals and data papers fulfil a similar role as they describe how datasets are collected, processed, and verified, helping thereby giving credit to data producers, and improving the control of data provenance [10]. Data papers contain facts about physical properties and precision of the data, as well as describing published research datasets and providing contextual information about their production, acquisition, and potential utility. Data papers are citable and provide crosslinks to data repositories. Together with data journals, they integrate open access journal platforms and data repositories into a complex academic structure [11].

### 2.2. Research ethics, data ethics

Researchers and all kinds of information professionals should be aware of the importance of ethical concerns. Research ethics is a unique part of professional ethics, for rightfully acquiring research funding, and meeting the conditions necessary for publication. In the same time, it is closely related to social responsibility. It should involve honesty, objectivity, morality, prudence, openness and respect for intellectual property, confidentiality, responsible management and publication, respect for colleagues, social responsibility, and the security of people involved in the research. Ethical approval to research must protect research participants' rights and welfare, and reduce the risk of physical and psychological discomfort, damage, and/or threats appearing because of research procedures. It should care for the rights of the researchers to carry out research in a lawful manner and care for the reputation of the institutions by excluding the negligence of researchers and cooperating persons [12].

Research ethics is closely tied to data ethics that is a "branch of ethics that studies and evaluates moral problems, related to data (including generation, recording, curation, processing, dissemination, sharing and use), algorithms (including AI, artificial agents, machine learning and robots) and corresponding practices (including responsible innovation, programming, hacking and professional codes), in order to formulate and support morally good solutions (e.g. right conducts or right values)." It is based on computer and information ethics, highlighting the need for ethical analyses to concentrate on the content and nature of computational operations, rather than on digital technologies [13]. The compliance with ethical rules is a prerequisite of a responsible use of data that equals to the ability of using appropriate data for appropriate purposes, giving attention to integrity, accountability, transparency, privacy, and confidentiality of data [14].

Research ethics is highly relevant for the evaluation, monitoring and governance of research activities in universities and research organizations. It is apparently becoming increasingly important at least on two levels:
1. Data models that enable representing ethical aspects of research projects.
2. Development, implementation and functioning, enabling compliance with usual (actual) ethical standards of scientific research [15].

## 2.3. The role of data professionals

There are different data professionals who have the potential to support the work of researchers. Many of their responsibilities are still evolving.

Data librarians, whose duties partially overlap with the ones of data curators and data stewards must be mentioned here, even if data curation is often regarded to be more domain-specific than data stewardship [16]. Data librarians must have a conceptual understanding of data, and be familiar with the surrounding issues, as well as being able to find, extract, collect, clean, organize, analyze, and present data. They may be able to be involved in data support, research data management, data curation, data governance, data quality evaluation, data citation, and fostering data literacy. The duties and skills required of them include being familiar with the related legal requirements and ethical considerations [17].

Data stewardship is also an important building block of proper data management infrastructures, thus this term is sometimes used interchangeably with data governance [18]. Its essence is "releasing some kinds of data and preventing the release of other kinds of data. The same data may fall into either category, depending on the time, purpose, or entity requesting access" [19].

Data stewards ensure that working with data is performed according to policies and practices as determined by data governance. From among their varied roles, data stewards insure informed decisions, based on quality data [20]. They represent the concerns of others, i.e. take care of data assets that do not belong to the stewards themselves. The latter can be defined as: "Organizations and their personnel defining, applying and monitoring the patterns of rules and authorities for directing the proper functioning of, and ensuring the accountability for, the entire life-cycle of data and algorithms within and across organizations" [21]. Data stewardship strategies include caring for data accessibility, ease of use by data providers, receivers and intermediaries, caring for flexibility, robustness, system availability and integrity, level of sophistication, integration and interoperability, as well as auditability [22].

## 3. Research Information Management: State of the art

Research Information Management (RIM) is a service that refers to the "management, evaluation, and disclosure of research outcomes and expertise, which connects in various ways with internal evaluation and management goals, funding policy and compliance needs, as well as with broader reputation management on the web" [23].

RIM is different from RDM, as it aims at "uniting information from the different institutional systems under a common interface as they allow an organization to achieve the full array of life-cycle needs, from the identification of funding, through reporting and benchmarking of research outputs, to high level strategic planning" [24].

Research Information Management Systems (RIMS), or Current Research Information Systems (CRIS) collect and store metadata on research activities and outputs, such as researchers and their affiliations; publications, datasets, and patents; grants and projects; academic service and honors; media reports and statements of impact [25]. RIM systems enable access to the identity information of researchers and the content, authored by them. Research identity may include information about publications, datasets, and research technologies, produced by researchers. It also provides information about the researcher's capabilities, skills, and expertise [26].

The emerging role of artificial intelligence analysis of research information with emerging topics like AI is one of the most promising scenarios for deployment in different institutions. Many academic institutions are now recognizing the value that their research information can represent. The generation of value from the available research information is one of the most important tasks of today's IT departments. Research information developments can only be as good as the quality of the research information resources allows. When supporting specific processes, high quality of data and information are required and must be continuously improved. This requires AI methods that can deal with today's complexity of data and information in an automated and self-learning manner.

By leveraging big research information, techniques can be used to address the challenges of dealing with data quality issues and bring greater value to higher education and research institutions. In addition, a survey showed that most universities are dissatisfied with the quality of research information collected and processed, rating it as low or fairly low. However, they often do not take any action to improve it [27].

In the 20th century, research into data quality problems such as anomaly detection in computer science and statistics began [28, 29]. Research has been done to detect outlier quality issues, which have been described in more detail in the literature. Since then, other techniques to detect data quality problems have been developed in various research and application areas [30]. Applying other techniques, methods, and algorithms can help identifying this quality, e.g. signal processing, data filtering, and data integration. There are statistical methods that characterize the data, deriving key figures or features, time series analysis. Moreover, we can count with methods, such as data visualization, data mining. Machine learning (ML) methods can be applied for cleaning the data (e.g. by replacing missing values, duplicate detection). We can also engage in actual modelling. This also enables databases and IT systems to use algorithms to recognize patterns and connections from existing data for example, predictions can be made and business processes can then be automated. With large amounts of data available, deep learning has also recently gained importance. This is based on algorithms that can recognize text and data patterns. In addition to data profiling and data cleansing, data integration and enrichment play an important role in merging often heterogeneous source data. In this context, semantic technologies, such as knowledge graphs are also used, the facts of which are used to validate, evaluate and annotate data, among other things. When using linked data, knowledge about individual data can be networked, enriched and new information gained from their combination.

The field of data engineering and data science deals specifically with these techniques, methods and algorithms for the collection, management, storage, processing, enrichment and provision of research information, which are therefore used below as a generic term for covering not only the concept of data management, but the integration and preparation of data. In addition, an analysis of the research information is essential to assess the quality of the data, to be able to interact with it effectively.

Artificial intelligence has attracted much attention from government, industry, and academia [31]. In the literature, many authors developed different quality models, using artificial intelligence algorithms like ML to evaluate the quality in order to analyze data in systems, but specify and evaluate them objectively [4, 32, 33].

There is little research on the application of AI to ensure data quality on RIM. This research is needed as AI application for decision-making is shifting from operational areas of RIM to strategic ones. Therefore, we must focus on new research topics such as AI and algorithms for machine learning in particular making use of artificial neural networks as an alternative to data quality methods such as data profiling, data cleansing and data wrangling [34, 35, 36] in the context of RIM. Assessing and improving the quality of research information using artificial intelligence offers many benefits, such as:
- Improving existing processes for data quality through automation,
- Errors can already be detected and corrected during loading,
- Early detection of data errors leads to time savings in data quality processes Easier control and transparency,
- Cost reduction by saving resources on repetitive tasks,
- Higher acceptance among employees.

## 4. AI Methods – How can patterns be identified?

Behind the practical applications of artificial intelligence are basic technologies, such as ML, computer vision or Natural Language Processing (NLP). They lay the foundation for what constitutes intelligent machines: the ability to recognize patterns in large amounts of data - and, even more importantly, to learn and optimize processes independently. However, there are also technical and statistical limitations that must be taken into account for a realistic assessment of the effective performance of artificial intelligence. Such an assessment is elementary in order to make strategic decisions that are successful in the long term with a view to the development and launch of new AI systems.

The basic ability to classify research information is the cornerstone of concept formation, abstraction and inductive reasoning and thus of human intelligence. It makes it possible to bring order to the often chaotic variety of sensory perceptions. This requirement also applies to artificially intelligent systems: A fundamental ability that characterizes artificial intelligence systems is the identification of structures through regularities and patterns within datasets.

To translate this capability into a benefit, research information must first be converted into a digital form so that RIM can process the data. Pattern recognition searches for statistical correlations, similarities, regularities or repetitions.

AI systems can simulate some level of language capability by using algorithms to identify rules within text files and apply them to enable interaction between humans, databases, and machines. In the interdisciplinary research field of NLP, linguistics works together with computer science on the analysis, interpretation and generation of natural language. Since spoken or written conversations and statements serve as a basis, the RIM use unstructured data in most cases. Various techniques and methods are combined to analyze language from research information to meaning. After the data has been converted to text form, the strings of letters can be broken down into word units using tokenization. There are various additional steps to syntactic processing, such as morphological segmentation, in which case markers, plural forms or word combinations are interpreted to derive grammatical information. Finally, the meanings and relationships between words and sentences can be mapped through semantic and dialogue analyses. Specific procedures for proper noun recognition can help to automate the classification of authors, affiliations or places. In the field of NLP, individual language components are analyzed and transferred to the overall language system. Speech recognition in communicative assistance systems such as Siri as the control software for Apple products or Alexa from Amazon for interaction with digital applications or the smart home infrastructure is based on these principles.

For the analysis of research information, text mining makes it possible to automatically work out meaning structures within unstructured text data. The method is particularly suitable for extracting relevant information and statements from large text files. Text mining makes it possible to generate hypotheses with the help of algorithms and to check them automatically.

In principle, all research information can be analyzed and classified by pattern recognition. The field of pattern recognition is therefore extremely broad and ranges from the analysis of voices and handwriting to the evaluation of text documents and video files for the detection of objects in road traffic or in space.

Probably the most important sub-discipline of artificial intelligence is machine learning that consists of processes and algorithms that can train themselves. The way it works goes well beyond conventional if-then operations: ML is real learning based on datasets. It can automatically generate knowledge, train algorithms, identify connections and recognize unknown patterns. These identified patterns and relationships can be applied to a new, unknown dataset in order to make predictions and optimize processes. This type of learning is often associated with the terms data mining and predictive analytics.

## 5. Successfully implementing AI methods in RIM

As soon as a use case, providing a written description of how users will perform particular tasks has been identified, a project team and a suitable infrastructure have been defined, the implementation of the application based on AI methods can begin. As a part of this paper, a process model and concepts are presented that displays the elementary phases from the start of the project to the ongoing operation of the AI methods in the RIM. The approach presented below (see Figure 1) is based on the standardized process model CRISP-DM (Cross Industry Standard Process for Data Mining), which was developed in 1996 by cross-industry partners [37]. The limited amount of data in AI projects compared to data mining results in extended requirements with regard to the models used. In order to meet these requirements for AI applications when setting up and maintaining the systems, the CRISP-DM model was adapted for the implementation of AI projects [38]. In order to successfully implement the AI methods, the project team goes through various project phases sequentially. Depending on the specific project, the various phases can also be combined, skipped and repeated in the sense of agile development. A uniform sequence of phases can therefore only be determined in advance to a limited extent.

Figure 1 illustrates eight different project phases, which are usually passed through from the start of the project to ongoing operation. Usually, phases 4 (modelling) and 5 (evaluation) are run through iteratively until the required model quality can be achieved.

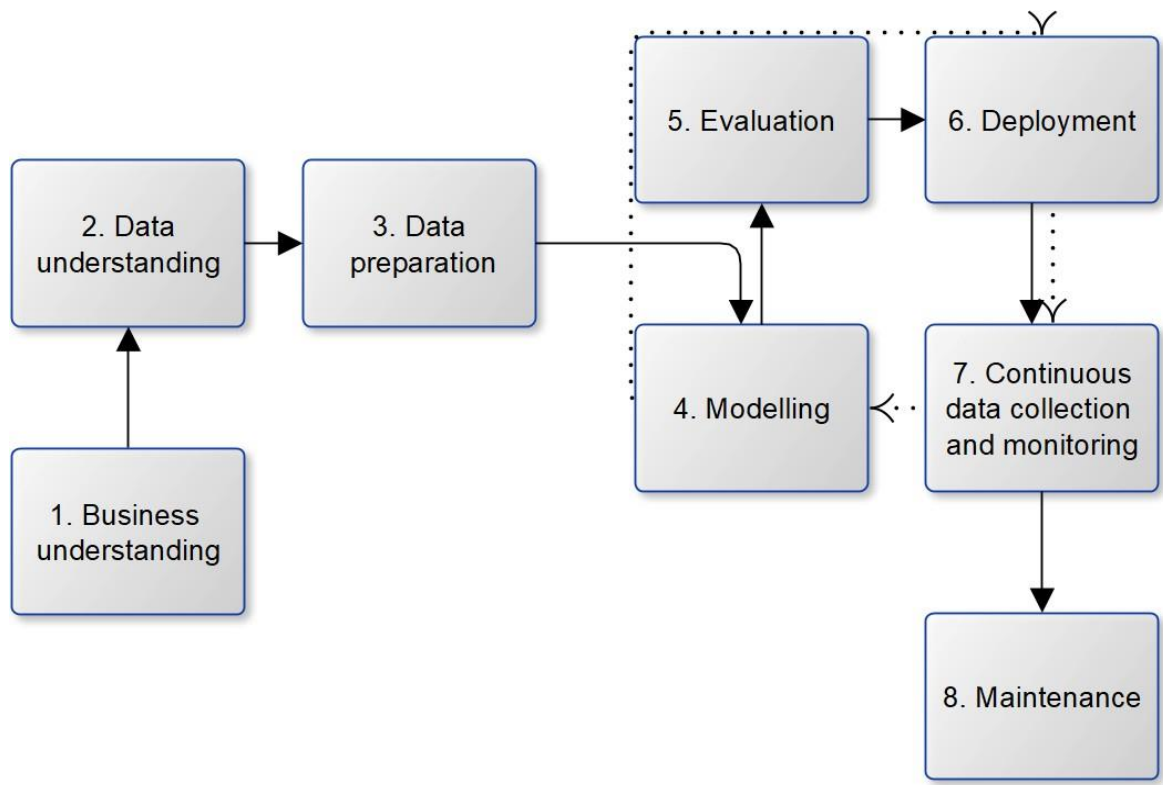

**Figure 1**: Procedure model for implementation of an AI project

The individual project phases are briefly explained below.

**Phase (1):** Once a suitable use case has been identified and the project is approved, a common understanding of the business and the project should be created for all, those involved. To do this, it is necessary to specify the specified requirements, goals, framework conditions and success criteria in the first project phase by working out a preliminary project plan together. Therefore, the following questions must be answered:
1. Which business areas are affected and are they involved?
2. Are all project participants aware of the available resources, such as personnel, software and hardware?
3. Who uses the AI methods and how should the AI be interact with?
4. Should only instructions be given, should the AI automatically adjust parameters or even intervene directly in the process control?
5. Are the terms defined across teams?
6. Has an initial project plan been jointly agreed?

**Phase (2):** After a common business and project understanding of all those, involved has been created in the previous step, the data, data structures and data sources that are already available are first collected, viewed and processed. The process and technology experts are significantly involved here due to their extensive knowledge and experience with the underlying process or the underlying technology. The aim of this second phase is a structured analysis of the available data and its origin. In the course of this analysis, it should be examined whether the data can be collected again in a later project phase with reasonable effort in order to retrain the AI methods in RIM. For this phase, the following questions must be answered:
1. Which data sources are available and are the datasets accessible and usable, taking data protection into account?
2. Which datasets can be taken from the available data sources?
3. In what form is this data available, e.g. as continuous process variables or as discrete variables?
4. What types of data are available, e.g. time series, spatial information or semantic relationships?
5. Which datasets are measured or manipulated?
6. Is the data collected continuously in a process under consideration or does it come from an external source?
7. Can data quality be evaluated already in this early project phase?
8. In what quantity are the retrospective datasets available?
9. Are there data gaps or frequent outliers?
10. If necessary, can the data also be recorded with little effort during subsequential operation?
11. Are all relevant process steps and influences visible in the data?

**Phase (3):** Before data-based modelling can be started, the data to be used must be selected, cleaned, filtered and formatted depending on the modelling goals. In many projects, this phase can take up a considerable, if not the largest, amount of time, which is why it should be given special consideration in project preparation and project planning. If necessary, meaningful characteristics for data-based modelling can be identified and constructed on the basis of the processed data. As an alternative to deriving characteristics, the subsequent modelling step can also be carried out directly with the prepared data is done. The basic decision as to how much technology and process knowledge is included in the so-called feature engineering (generation of features from the data) depends on how much data is available and how much the technology and process understanding can be broken down into individual features or how much technology and process knowledge should be integrated into the AI system. This phase of feature identification, extraction and construction is therefore often an elementary building block in the development of an AI application. The following questions must be answered for this phase:
1. Is the data sufficiently filtered and prepared, for example with regard to outliers etc.?
2. Are there similar problems that have already been solved with the available data? If so, which features were used?
3. Which physically motivated features can be identified from the data?
4. What artificial features can be generated from the data?
5. Can a connection already be identified between the output variables of the AI to be modelled and the possible characteristics?

6. To what extent does the choice of characteristics restrict the selection of possible modelling approaches in this phase?

**Phase (4):** After an understanding of the data and processes have been created and the data has been prepared and cleaned, this phase is about selecting the appropriate algorithms and creating models in order to be able to derive the desired knowledge gain from the collected data. There are a number of algorithms, suitable for creating models. The aim of this fourth phase is to select the appropriate modelling techniques, to create a test model and to evaluate it with regard to the prediction quality. The key stakeholders here are the AI experts who work together to select the parameters for modelling and evaluation. Often the parameters and input data are adjusted at this stage as a result of the evaluation and multiple models are built. This phase is therefore directly related to the previous one and can also be run through iteratively. The more clearly the characteristics and relationships of data can be defined and specified, the easier it is for the model to process such information in large quantities and deliver a reliable result. However, collecting data and training the algorithmic models is time-consuming and expensive. Therefore, it is recommended to start with simple models to create a reference for model evaluation. For the fourth phase, the following questions must be considered:

1. What input and output data does the model have?
2. Which algorithms are available for the given input data and the required output variables? What are the pros and cons of each?
3. Does the model have to be explainable?
4. How safety-relevant are the model decisions?
5. For which range of the variable to be modelled is a reliable result required? How reliable does the model have to be?
6. How can the reliability of the model be tested?
7. Is the model static or should it continuously adapt to new circumstances?
8. Can simulations be used to train the model?
9. How limited is the modelling when considering privacy? Does data have to be anonymized? Can sufficient accuracy be achieved without using the sensitive data? Is it permissible to use cloud solutions for the data?

**Phase (5):** In this phase, one or more models were created and evaluated from a statistical point of view. In the evaluation phase, the decision should be made as to whether the selected model meets the success criteria and requirements defined in the first phase, or whether a new loop should be drawn up. Before deploying the model, it is important to evaluate it and determine whether it can achieve business goals in the field and compensate for the impact of mostly unavoidable false predictions. In addition, each step in creating the model is checked before the model is subsequently released for use in the company. Since practical aspects play an important role here, the involvement of the people responsible for the business is essential in addition to the AI experts. The following questions must be answered for the evaluation phase:

1. Can we prove expect added value by piloting?
2. What could be the impact on the facilities of an erroneous output from the model? Can the objective be achieved by using the model?
3. Are further iterations and adjustments of the previous phases necessary?

**Phase (6):** After the release of the tested models for the operational deployment, the aim of this phase is to provide a user-friendly and operational solution. Even if the solution is essentially based on methods from the field of AI, existing control and regulation systems should not necessarily be substituted, but rather the AI approaches should be embedded in existing systems and platforms. Central to the implementation are questions about the function as well as the operation and intervention by the users. Expertise in the areas of software development and system administration is essential for this, which must be supplemented by the technological understanding of experts in order to identify domain-specific limits for the AI and RIM system and to be able to design a secure framework for operation. The following questions must be clarified:

1. How must the product be monitored in use?
2. In which situations is the RIM system and AI used?
3. Is it a stationary or a learning solution?
4. How do users need to be trained?
5. How do users report errors?

6. Does the RIM system have to be able to be retrained with AI applications?
7. How must the input data be monitored to ensure that the model does not extrapolate or encounter anomalies that the model is not aware of?
8. In which existing infrastructure will the RIM system and AI methods be integrated?

**Phase (7):** The dotted circle in Figure 1 describes the evolution of the model after it was first introduced in the field. The important step in the figure is the collection of data. Based on the feedback from the field and the insights gained from the model, further parameters and possible data sources can be identified that can supplement the model. This seventh phase involves the continuous collection of data and monitoring for additional relevant data to increase the accuracy of the model or to refine it as necessary. The domain and AI experts are responsible for these tasks as the main participants. Therefore, the following questions must be answered:

1. Is there a new data source or new parameters to be integrated into the model?
2. How to collect the necessary data to feed it into the dynamic model?
3. In what time interval should the monitoring of the data and its collection take place?
4. Has the data changed structurally compared to the data, initially used?

**Phase (8):** According to the design of the function, the operation and the possibilities for intervention by the users from phase 6, the development is monitored over the entire period of use in the maintenance phase. At the end of this phase, the result is a maintenance plan that shows how safe operating conditions can be guaranteed at all times. In addition, the following questions must be clarified:

1. Which critical situations can the RIM system with AI applications be exposed to due to a change in the input data in long-term use?
2. How and by whom are faults reported by users rectified?
3. Can the AI be spot-checked or does it have to be continuously monitored?
4. How can the cycles for retraining the models (maintenance plan) be determined?
5. How can it be ensured that the model in operation reacts sufficiently effectively and efficiently to the current input data?

The main phases represent the static project process and the most important activities per phase. The aim of the AI project presented, was to automate a visual inspection process by using AI in RIM systems in order to relieve RIM users and employees in facilities of the visual inspection and to establish a correlation between process parameters and component quality in the future.

The success of projects in the field of AI is often difficult to assess in advance, especially if there is no empirical value to assess the database and the chosen procedure for the individual application. For AI projects, where there is uncertainty about the information that can be extracted from the data and the knowledge that can be derived from it, it is advisable to demonstrate its feasibility as quickly as possible or to abandon the project if the data basis is insufficient. It is often domain or method-specific challenges and subtleties that decide on the feasibility of data-driven solutions.

So that AI-based solutions become unified in the long term can make a positive contribution to the profitability of RIM use in institutions, whereby it is of crucial importance to prepare documentation carefully in order to provide both successful and unsuccessful approaches. In this way, the specific challenges and benefits can be assessed afterwards and the knowledge gained can be used for subsequent AI projects. Based on the project documentation, the following key questions can be answered:

- What specific benefits can be achieved through the AI approach?
- What specific problem was solved?
- What specific challenges did you have to overcome to achieve this?
- How can good decisions made with the help of AI?

## 6. Discussion

Research information and data form the basis for all RIM and AI systems, which use ML methods to recognize the laws, hidden in the existing data and use them to derive information from data obtained later. The availability, quantity and quality of the data therefore play a decisive role in terms of the feasibility and results of an AI project.

An exclusion criterion for every AI project is therefore the lack of availability and applicability of existing data. Even if research information has already been collected, it may not be accessible because, for example, researchers do not want to release the data or personal data, or may only be used in compliance with legal requirements. In addition, data from different sources is often used in an AI project. If the data cannot be assigned, it cannot be used in such projects, therefore it is not feasible due to a lack of data availability.

The amount and quality of research information and data are the decisive success factors for the results of an AI project and the usability of the targeted AI system. Therefore, domain experts should assess whether the available research information is representative of the process under consideration and whether it contains the phenomena that can occur later in the operation of the AI system. For example, an AI system for predicting data quality (predictive quality) can only identify data errors from research information that are contained in the dataset under consideration with similar system states. Even when this requirement is met, a low number of data points may be available to learn the underlying laws. The more complex the relationships are, the more data points are usually required. Collected research information also often contains errors, such as incorrect or missing values. In addition, human errors cannot always be avoided when manually recording research information, and technical errors can also occur during data recording. Such errors can be dealt with through increased effort within the framework of the AI project. However, if there are too many errors, this can severely impair the performance of the AI system. A meaningful assessment of whether the data volume and data quality are sufficient for the use of the AI methods in RIM can usually only be made during implementation. Then it becomes apparent whether the research information can be used to derive generalizable principles for use in later operations. There are also a number of approaches to deal with small amounts of data. In the field of image recognition, for example, the amount of image research information can be artificially increased by rotating, shifting and scaling (by making use of so-called data augmentation in RIM). Probabilistic methods can also be used if the amount of data is not representative enough. In the field of RIM, the handling of small amounts of data is currently a topic that is dealt with intensively, with new methods being continuously developed, such as speech and image recognition with only a few speech and image recordings or the transfer of learned relationships over several tasks.

In addition to the technical limits, legal and ethical framework conditions should also be considered. As mentioned in the context of data availability, personal data (e.g. age, photos, weight) can be used in AI applications, whereby the General Data Protection Regulation (GDPR, https://gdpr.eu/) must be taken into account and appropriate measures must be taken to meet the legal requirements (e.g. creation and collection of data protection declarations). In particular, those data that can be used for a personal performance evaluation must be viewed critically. In addition, AI applications are used to make or prepare decisions that can influence people (e.g. in personal decisions) and should therefore be easy to understand. Such requirements for the transparency of decision-making must be defined at the beginning of a project, since the selection of the AI methods used, for example, can depend on this.

## 7. Conclusion

In today's data-driven research environment the value and importance of research information and professional RIM is undeniable, especially if there is proper planning and the systems are implemented properly. In treating the related issues, distinguished attention was given to the quality of data. Using an interdisciplinary approach, this paper argued for making use not only of approaches and tools, driven by artificial intelligence, but taking account of traditional views on data ecologies, embodied in information ecology, Research Data Management (Research Data Services), the roles of varied data professionals, as well as research ethics and data ethics.